%% file: main.tex
\documentclass[aps,prb,twocolumn,groupedaddress,showpacs,amsmath,amssymb,amsfonts,10pt]{revtex4-2}
\include{preamble.tex}

\begin{document}

\title{Cognitive Energy Cost of Informed Decisions}

\author{Michele Vodret}

\email{mvodret@gmail.com}

\date{\today}

\affiliation{\vspace{0.2cm}Université Paris-Saclay, CentraleSupélec,
Laboratoire de Mathématiques et Informatique
pour la Complexité et les Systèmes,
91192 Gif-sur-Yvette, France. \vspace{0.2cm}}

\begin{abstract}
Time irreversibility in neuronal dynamics has recently been demonstrated to correlate with various indicators of cognitive effort in living systems. Using Landauer's principle, which posits that time-irreversible information processing consumes energy, we establish a thermodynamically consistent measure of cognitive energy cost associated with belief dynamics. We utilize this concept to analyze a two-armed bandit game, a standard decision-making framework under uncertainty, considering exploitation, finite memory, and concurrent allocation to both game options or ‘arms’. Through exploitative, prediction-error-based belief dynamics, the decision maker incurs a cognitive energy cost. Initially, we observe the rise of dissipative structures in the steady state of the belief space due to time-reversal symmetry breaking at intermediate exploitative levels. To delve deeper into the belief dynamics, we liken it to the behavior of an active particle subjected to state-dependent noise. This analogy enables us to relate emergent risk aversion to standard thermophoresis, connecting two apparently unrelated concepts. Finally, we numerically compute the time irreversibility of belief dynamics in the steady state, revealing a strong correlation between elevated - yet optimized - cognitive energy cost and optimal decision-making outcomes. This correlation suggests a mechanism for the evolution of living systems towards maximally out-of-equilibrium structures.
\end{abstract}

\maketitle


\thispagestyle{empty}
\section{Introduction}

\label{sec:0}

Decision-making is a universal cognitive process~\cite{rangel2008framework}, manifesting across the entire spectrum of life as we understand it. This process requires a careful balance between exploring new opportunities and exploiting existing knowledge.

Significantly, exploitative behavior leads to time irreversibility. An action is considered irreversible if it notably reduces the range of future choices for an extended duration~\cite{henry1974investment}. Recent advancements in the neural foundations of decision-making~\cite{padoa2017orbitofrontal} inspire our exploration of time irreversibility within belief dynamics.

To distinguish time irreversibility in action dynamics from that in belief dynamics, consider a scenario where an individual allocates limited resources between two options, $A$ and $B$. Unbeknownst to the individual, both options offer unknown but statistically equivalent rewards. An initial preference for $A$ over $B$ paves the way for exploitation. Depending on the exploitation intensity, belief dynamics might demonstrate a cycle of self-fulfilling prophecies: a bias towards option $A$ increases resource allocation to it, resulting in higher average rewards and reinforcing the initial bias. This cycle persists until negative fluctuations in the favored option shift preference to the other. Over time, though belief dynamics are time-irreversible due to resource-limited exploitation, the resultant resource allocation and acquired rewards could display time-reversible dynamics. We will formalize this observation using a stylized decision-making model.

The aforementioned self-fulfilling prophecy mechanism plays a crucial role in various social contexts, encompassing financial markets~\cite{marsili2001market,wyart2007self,vodret2023microfounding} and economics~\cite{farmer1999macroeconomics,bouchaud2023self}, information dissemination in social media~\cite{marsili2004rise,da2015sudden,cinelli2021echo,mocanu2015collective}, the dynamics of politicians and voters in election polls~\cite{frisell2009theory,rothschild2014public}, up to war engagements scenarios~\cite{merton1948self}. This highlights the importance of studying single-individual belief dynamics in order to understand how collective behaviors emerge.

We have chosen model-free Reinforcement Learning (RL)~\cite{daw2014advanced} for our case study, capturing how subjective values or beliefs for each option are independently assessed and incorporated, adapting to novel opportunities. Instead of constructing an environmental model to optimize each action towards a set goal, this class of algorithms directly determines the subjective value functions from interactions with the environment. One prominent algorithm in model-free RL is Q-learning, which, in its fundamental form, updates the subjective value of available options based on prediction errors. Notably, this framework has been employed recently to model human cognitive biases, such as positivity or confirmation bias~\cite{palminteri2022computational,palminteri2022choice}, in two-armed bandit tasks. 

We investigate the influence of time irreversibility tied to exploitative behavior stemming from prediction-error-based belief dynamics in a two-armed bandit problem.
Our decision-making model links time irreversibility in belief dynamics to a thermodynamically sound concept of cognitive energy cost via Landauer's principle~\cite{landauer1961irreversibility, frank2018physical, berut2012experimental}: time-irreversible information processing generates heat. Recent discussions have considered time irreversibility at the neuronal level, revealing a significant correlation between established cognitive effort proxies and irreversibility in fMRI and MEG human-brain data across a variety of tasks and conditions~\cite{lynn2021broken,perl2021nonequilibrium,deco2022insideout,gilson2023entropy,tewarie2023non,bernardi2023time}. Our contribution focuses on the more abstract belief space, leading to the cognitive energy cost concept.

 This study offers three primary takeaways: $i)$ A formal merging of emerging risk-aversion and thermophoresis - the tendency of solute particles to migrate towards cooler regions. $ii)$ A connection between time irreversibility of intertwined belief dynamics, dissipated work, and cognitive energy cost. $iii)$ From a comprehensive theoretical and numerical analysis, we discern that intermediate exploitative behavior aligns with a peak, yet optimized in a precise thermodynamic sense, cognitive energy cost, and an effective balance between exploration and exploitation.

The following sections are structured as follows for the 
reader's ease: section~\ref{sec:1} presents a modified version of the forgetting Q-learning model. Section~\ref{sec:Gammas_behavior} explores the relationship between exploitation and time irreversibility in belief dynamics using a spatially coarse-grained description. Section~\ref{sec:fokker_planck} outlines the mapping of belief dynamics to an active particle model and discusses the link between emerging risk aversion and thermophoresis. Section~\ref{sec:methods} initially delves into the general association between time irreversibility in belief dynamics and cognitive energy cost and later shares numerical findings related to the modified forgetting Q-learning model. Section~\ref{sec:3} concludes the discussion and suggests potential avenues for future research.

\section{Forgetting Q-learning model with concurrent investment}
\label{sec:1}

Consider a two-armed bandit game scenario where, at every time step \( t \), a decision maker has to invest a single unit of endowment between two `arms' of a slot machine, denoted as $A$ and $B$. 
$a_t \in [0,1]$  signifies the investment fraction at time step \( t \) on bandit \( A \), while \( 1-a_t \) does so for bandit \( B \). 

    The rewards yielded by the arms at each time step are \( R^A_t a_t\) and \( R^B_t (1-a_t)\), respectively. Both $R^A_t$ and $R^B_t$ are drawn by time-independent Gaussian distributions, chosen arbitrarily such that the support is mostly in the interval $[0,1]$. We indicate the mean and variance of $R^A_t$ respectively as $\langle R^A\rangle$ and $\sigma^2_A$, with analogous notation for $R^B_t$; these pieces of information are unknown to the decision maker.

     A natural choice~\cite{palminteri2022choice} is to let $a_t$ depend solely on the difference of the beliefs at the current time step~\( t \), denoted respectively as \( \hat R^A_t \) and \( \hat R^B_t \). A possible parametrization of $a_t$ is
    \begin{equation}
    \label{eq:meana}
         a_t  = \cfrac{1+\tanh\left[\Gamma (\hat R^A_t-\hat R^B_t)\right]}{2},
    \end{equation}
    where $\Gamma \geq 0 $ is the exploitation parameter: it dictates how belief disparities affect investments. A positive exploitation parameter \( \Gamma \) value enhances the inclination to invest more in the arm perceived as more lucrative.

 The forgetting Q-learning model~\cite{Katahira,seidenbecher2020reward} prescribes the following belief dynamics:

\begin{subequations}
\label{eq:RW+f}
\begin{align}
\label{eq:RW+f_A}
\hat R^A_{t+1} &= \hat R^A_t + \beta a_t (R^A_t-\hat R^A_t)-\beta (1-a_t)\hat R^A_t,  \\
\label{eq:RW+f_B}
\hat R^B_{t+1} &=  \hat R^B_t + \beta (1-a_t) (R^B_t-\hat R^B_t)-\beta a_t \hat R^B_t. 
\end{align}
\end{subequations}
Here $\beta > 0$ manages two facets: the agent's sensitivity to new data via the prediction error and the agent's propensity to forget, which are embodied in the second and third terms of the equations above, respectively. Notably, the forgetting terms shift the beliefs towards zero, implying the agent takes the minimum reward (in this case, zero) as a reference. A byproduct of this choice is that, even for symmetric bandits, the agent might have a prolonged preference for one arm, leading to the emergence of effective trapping beliefs states.

\begin{figure}
        \centering
        \hspace{-9cm}\includegraphics[scale = 0.38]{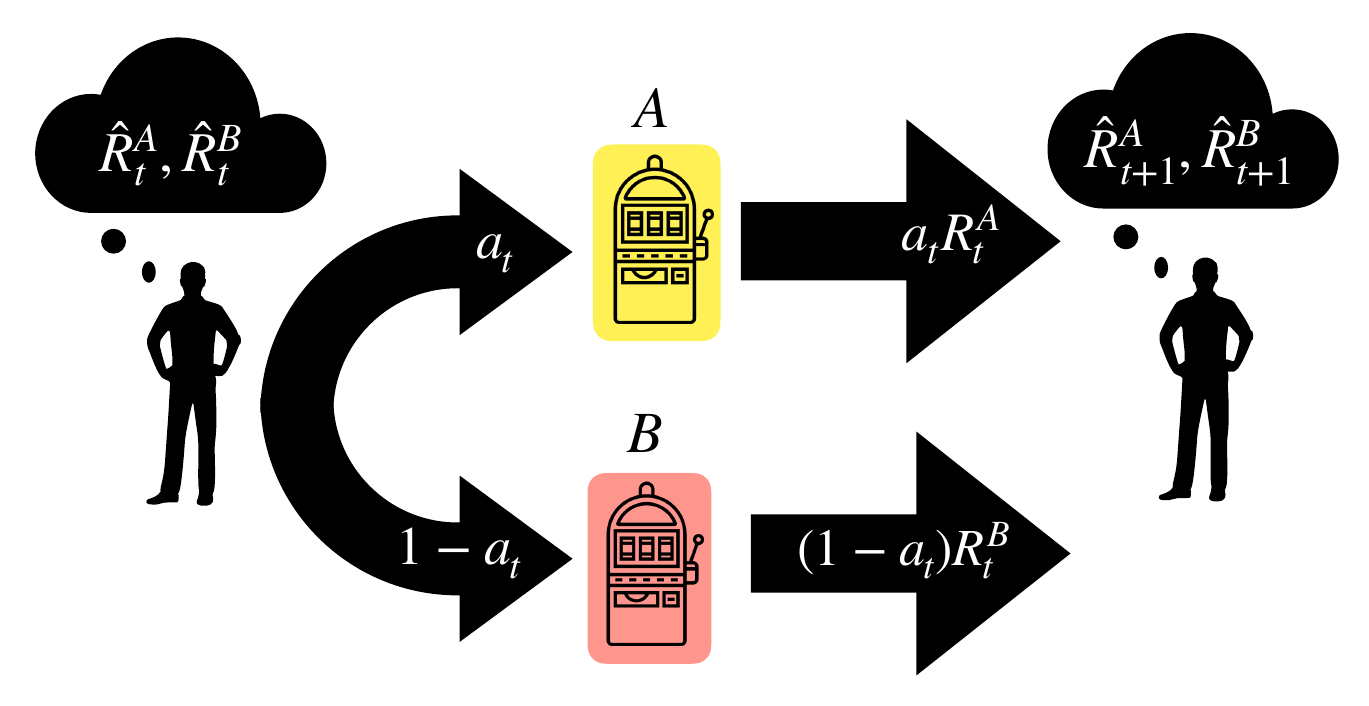}
        \caption{Graphical representation of the model. The investment decision precedes the observation of the actual outcome. }
        \label{fig:model}
    \end{figure}

 A graphical description of the dynamics between time step~$t$ and $t+1$ is given in Fig.~\ref{fig:model}: the action taken by the agent at time step $t$ is a function only of the current beliefs. Then, based on the obtained rewards, the agent updates his beliefs. Note that the beliefs dynamics is Markovian, i.e., the updated beliefs depend only on the previous ones.

 Before delving into the analysis of the dynamics of the proposed model, we shall address potential criticisms.

\subsection{Rationale for modifications}
\label{subse:modification}

We modified the forgetting Q-learning model in two ways with respect to the one discussed in the existing literature~\cite{Katahira}.

First, we consider $a_t$ to be a deterministic variable, while customarily it is distributed accordingly to a Bernoulli distribution; this is not a crucial characteristic of the model and, since it introduces a third source of noise, we neglect fluctuations of this variable. In doing so, we let $a_t$ vary continuously, incorporating a notion of confidence~\cite{confidence} in the model. A rationale for this choice is to consider $a_t$ as a time average along dynamics where the state, i.e., the couple of beliefs,  changes slowly; in this case, fluctuations are averaged naturally and the present choice is consistent with that presented in the available literature.

Second, using Gaussian rewards ($ R^A_t$ and $ R^B_t$)
  diverges from typical cognitive neuroscience models, where these are drawn from Bernoulli distributions~\cite{palminteri2022computational,palminteri2022choice}. This choice might be favored as it allows point estimates in the update equation to capture all noise statistics. 
However, we posit that Gaussian noise is more appropriate for the following reason: in passive learning scenarios ($\Gamma = 0$), where the investment is split equally on both arms ($a_t = 1/2$), the belief dynamics given by Eqs.~\eqref{eq:RW+f} result in two independent first-order Auto-Regressive (AR) processes. These are not time-reversible in the long run~\cite{weiss1975time} if Bernoulli rewards are considered. This contrasts with the time reversibility seen in Alzheimer’s Disease-related brain dynamics~\cite{cruzat2023temporal}. Moreover, it contrasts with a plausible application of Landauer's principle: passive learning in stable environments shouldn't expend cognitive energy. 
A way out is to consider Gaussian rewards, which give rise to time-reversible AR processes of the first order in the long-time limit of passive beliefs dynamics. A potential rationale for this is that learners make small errors when evaluating the update equations given by Eqs.~\eqref{eq:RW+f}, which effectively act as a form of spatial coarse-graining, effectively restoring time reversibility~\cite{busiello2019entropy} in the steady state.

\begin{figure*}[t!]
    \centering
    \hspace{-14.6
    cm}\includegraphics[scale = 0.45]{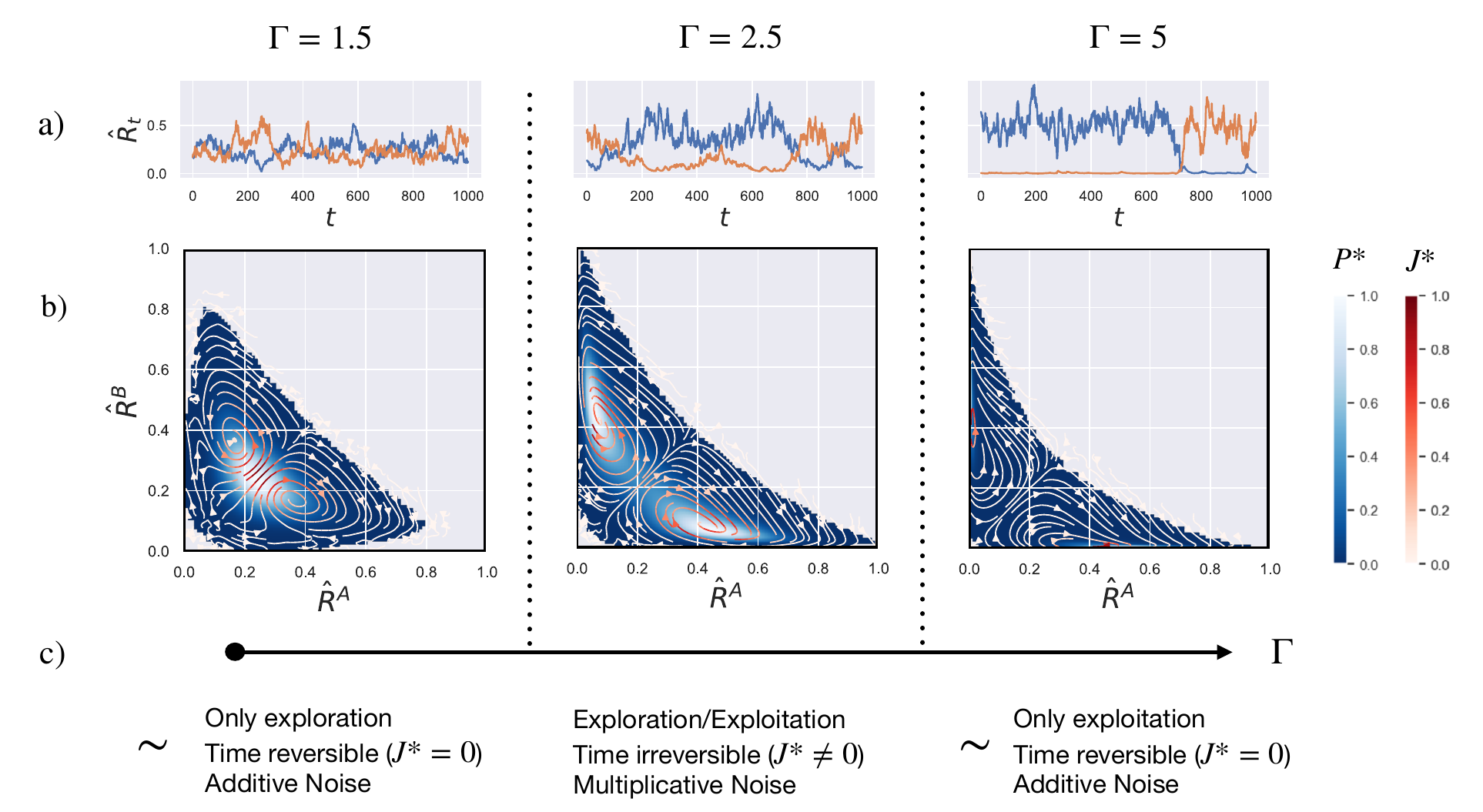}
    \caption{ Analysis of steady-state beliefs dynamics in the case of symmetric arms: $\langle R^A \rangle =\langle R^B \rangle = 0.5$, $\sigma^2_A = \sigma^2_B = \langle R^B \rangle (1- \langle R^B \rangle)$ and $\beta = 0.1$. $\mathrm{a})$ Belief dynamics for $\Gamma = 1.5$ (left), $\Gamma = 2.5$ (center) and $\Gamma = 5$ (right). $\mathrm{b})$ Steady-state. probability distribution on coarse-grained state space ${P_*}[\hat R]$, represented by the blue/white background color, together with stationary probability currents among coarse-grained states $J_*$ shown by red/white arrows. The scale of the color bars is purely qualitative.  $\mathrm{c})$ Main characteristics of the three phases of the model for increasing exploitative behavior. }
    \label{fig:regions}
\end{figure*}

\section{Preliminary analysis}
\label{sec:Gammas_behavior}
 Based on the exploitation parameter 
$\Gamma$, the belief dynamics provided by Eq.~\eqref{eq:RW+f} displays three different regimes detailed below.

\begin{itemize}[leftmargin=*,align=left]
    \item[\( \Gamma = 0 \):] the beliefs dynamics is passive, in the sense that the agent's action is decoupled from his own beliefs. In particular, the agent will always split the investment equally. Another way of formulating this concept is by saying that in the case of passive learning, there is no feedback between actions and beliefs. Therefore, the dynamics of the beliefs are completely decoupled and they evolve in time as independent first-order AR processes with Gaussian noise. Note that, as we stressed in Sec.~\ref{subse:modification}, the belief dynamics is time-reversible in the steady state~\cite{weiss1975time}.
    \item[$\Gamma \sim 1$:] investments influence the belief dynamics. This is because  
     $a_t$ is determined by the difference in beliefs, introducing a state-dependent, i.e., multiplicative, noise. The agent will mostly invest in the arm with the highest expected reward at the current time step. This region is the most interesting for us; let us mention here two reasons why: first, it is with a $\Gamma$ in this region that the learner will gain the most on average~\cite{palminteri2022computational,palminteri2022choice} in cases where $\langle R^A \rangle \neq \langle R^B \rangle$. Second, as I will show later in a precise sense, in this region the belief dynamics is time-irreversible even in the steady state. An intuitive understanding is the following: exploitation of past information leads naturally to an arrow of time.

        \item[\( \Gamma \gg 1\) :] in this situation one of the two beliefs is pushed to zero by the tendency to forget, therefore $a_t \sim  1$ or $a_t \sim 0$ for extended time periods, even if it is suboptimal. Effective trapping states, therefore, emerge, in which the beliefs are stuck, leading effectively to additive noise terms in the belief dynamics. In this scenario, the 2-dimensional stationary belief dynamics happens only in a 1-dimensional space, since one of the two beliefs is effectively frozen. The belief dynamics for large $\Gamma$s is therefore analogous to a single first-order auto-regressive process with Gaussian noise, recovering time reversibility in the steady state.
\end{itemize}
Refer to the panel $\mathrm{a})$ in Fig.~\ref{fig:regions} for an indicative example of belief dynamics $\hat R_t = (\hat R^A_t,\hat R^B_t)$ in these three regions. 

\subsection{Coarse grained analysis}
Alongside the visual inspection of the trajectories of the beliefs, an object worth analyzing is the probability density function (PDF) of the beliefs indicated as ${P}_t = {P}_t[\hat R_t]$, which offers a clear graphical picture of the emergence of trapping states. 

A spatially coarse-grained version of it is shown for the steady state of the system for different $\Gamma$s in panel $\mathrm{b})$ of Fig.~\ref{fig:regions}; note that there and in the following we will identify steady-state properties by the subscript~$_*$. One clearly observes the transition to bi-modality of ${P}_*$ as $\Gamma$ increases, related to the emergence of trapping states. Most importantly for the remainder of the paper, for moderate $\Gamma$ values, ${P}_*$ spans the 2-dimensional space maximally, while for large $\Gamma$s the beliefs dynamics is mostly constrained onto a 1-dimensional space. 

\setcounter{equation}{2}

$P_*[\hat R]$ alone does not give insight into the microscopic dynamics. In order to do that, a first approximation is given by Markov Chains. To monitor net movements for the spatially coarse-grained picture of the model one can compute the transition matrix $\mathcal{T}[\hat R^{i}\rightarrow \hat R^{j}]$ from state $\hat R^{i}$ to state $\hat R^j$, where $i,j \in \{0,\dots,N\}$, $N^2$ is the cardinality of the coarse-grained state space and $\mathcal{T}[\hat R^{i}\rightarrow \hat R^{j}]$ represents the probability of going to the coarse-grained state $\hat R^{j}$ starting from $\hat R^i$. From the transition matrix, one can define the associated probability current as
\begin{equation}
\begin{split}
\label{eq:cg_J}
    J_t[\hat R^{i}\to \hat R^{j}] =& {P}_t[\hat R^{i}] \  \mathcal{T}[\hat R^{i}\to \hat R^{j}] 
    \\
    &- {P}_t[\hat R^{j}] \  \mathcal{T}[\hat R^{j}\to \hat R^{i}].
\end{split}
\end{equation}

A useful classification of the system's dynamical state in the steady state is contingent on the value of the probability current:
\begin{itemize}[leftmargin=*,align=left]
    \item[$J_* = 0:$] In equilibrium steady states  (ESS)~\cite{gardiner1985handbook} there are no probability currents. This is indicative of time-reversal symmetry (TRS), i.e., in these states there is a complete absence of net movements in the system; this condition is known in the physics literature as detailed balance. In the dynamics of interest here, ESSs are observed in two distinct regimes of the exploitation parameter: \( \Gamma = 0 \) and \( \Gamma \gg 1 \). 
    \item[$J_*\neq 0:$] non-equilibrium steady states (NESS) exhibit probability currents. In systems with a compact state space, these flows lead to net circulating movements in the belief space, showing time-reversal symmetry breaking (TRSB).  This behavior is notably prevalent in the regime \( \Gamma\sim1 \) of the dynamics of interest here.
\end{itemize}

 Probability currents among coarse-grained states are depicted on top of the plots in panel $\mathrm{b})$ of Fig.~\ref{fig:regions}. 
 As can be visually appreciated, the NESS for $\Gamma \neq 0$ leads to a structure of probability currents similar to dipole currents~\cite{mendler2020predicting,busiello2023emergent}. 
The rise and fall of self-fulfilling prophecies anticipated in section~\ref{sec:0} is now evident: a small initial bias towards arm $A$ with respect to the equilibrium condition of passive learning ($\hat R^A = \hat R^B = 0.25$) leads to an increase in the value of $\hat R^A$ and a decrease of $\hat R^B$; eventually, $\hat R^A_t$  reaches the bottom right angle of the belief space and from there $\hat R^A$ will diminish; when $\hat R^A \sim \hat R^B$ two things can happen: or the initial bias is restored, and the cycle repeats itself, or there is an inversion such that $\hat R^B>\hat R^A$. In this latter case, $\hat R_t$ will follow the cycle in the upper triangle of the plot, completely analogous to the cycle in the lower triangle of the plot.

Panel 
$\mathrm{c})$ in Fig.~\ref{fig:regions} summarizes the three relevant $\Gamma$-dependent region of the present model.

\section{Continuous-time description}
\label{sec:fokker_planck}
In the previous section, we estimated currents between coarse-grained states. It is well known that the estimation of the probability currents $J$ on a spatially coarse-grained version of the system's state space provides only lower bound estimates on these~\cite{busiello2019entropy}. In order to properly estimate probability density currents, and therefore -as we will see in Sec.~\ref{sec:methods}- time irreversibility,  in a continuous-state system, a useful framework is given by Fokker-Planck equations; the reason for this is related to a technical simplification: the Fokker-Planck equation associated with a stochastic process is the deterministic dynamic equation for its PDF.

To this end, let us first perform the mapping from the belief dynamics to the corresponding continuous-time limit, from which the Fokker-Planck equation follows.

\subsection{Langevin equations}
\setcounter{equation}{3}

 For $\beta \ll 1$, the continuous time limit of  Eqs.~\eqref{eq:RW+f} reads
\begin{subequations}
\label{eq:cont_lim}
\begin{align}
\cfrac{d \hat R^A_{t}}{dt} &= -\beta  \hat R^A_t +  \beta a_t R^A_t,
\\
\cfrac{d \hat R^B_{t}}{dt} &=  -\beta \hat R^B_t + \beta (1-a_t) R^B_t.
\end{align}    
\end{subequations}
The coupled Langevin equations~\cite{gardiner1985handbook} above articulate how beliefs evolve in time due to drifts -or systematic tendencies- and diffusions, which refer to random fluctuation; the former is represented in our system by the forgetting term and the average noise-related contributions, while the latter relates to the deviation from the mean of the noise term in our model.

It is easy to focus directly on these two objects by rewriting Eqs.~\eqref{eq:cont_lim} in a more compact form given by  
\begin{equation}
\label{eq:langevin}
    \cfrac{d\hat R_t}{dt} = \mathcal{F}_t+\xi_t, \quad  \text{with}, \quad \langle \xi^T_t \xi_{t'} \rangle = 2 \mathcal{D}_t \delta_{t-t'},
\end{equation}
where  $\mathcal{F}_t$ is the drift vector, and $\mathcal{D}_t$ is the diffusion matrix, both of which depend on the current belief $\hat R_t$.

The mention of an important technical subtlety is mandatory here:
stochastic differential equations with multiplicative noise, such as Eqs.~\eqref{eq:cont_lim} or Eq.~\eqref{eq:langevin} for $\Gamma \neq 0$, require an interpretative framework, like Ito or Stratonovich~\cite{cugliandolo2017rules}, for concrete predictions. Due to the nature of the modified forgetting Q-learning model we are analyzing, where $a_t$ precedes the reward observation (see Fig.~\ref{fig:model}), the Ito interpretation is apt.

\subsection{Fokker Planck equation}
\label{sec:FP}
The Fokker Planck equation describes the deterministic dynamics of the PDF~${P}_t = {P}_t(\hat R)$ as 
\begin{equation}
\label{eq:FP}
    \cfrac{\partial {P}_t}{\partial t} = -\nabla \cdot J_t,
\end{equation}
where the probability density current $J_t$ is given by
\begin{equation}
\label{eq:current}
    J_t = \mathcal{F}_t{P}_t-\nabla \left( \mathcal{D}_t {P}_t\right).
\end{equation}  
 \( J_t \) represents here the local net flow of probability in the beliefs space $\hat R$ and it is the continuous -in time and space- analog of the probability current introduced in Eq.~\eqref{eq:cg_J} in the spatially coarse-grained description of the discrete-time belief dynamics.

\begin{figure}
    \centering
    \hspace{-9cm}\includegraphics[scale = 0.3]{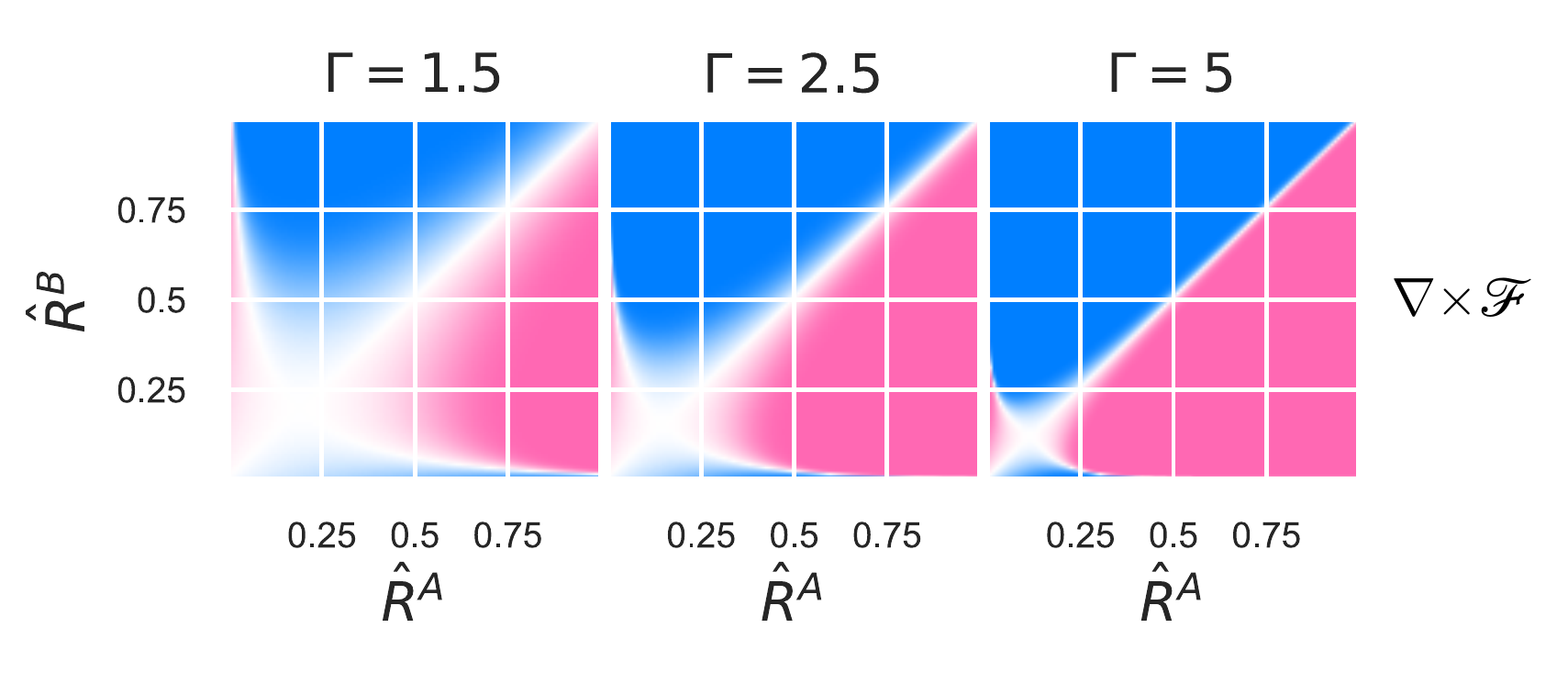}
    \caption{Curl of the thermodynamic force. The color ranges from azure ($+1$), to pink ($-1$), going through white ($0$). The plot is out of scale. The parameters are the same as Fig.~\ref{fig:regions}}
    \label{fig:curl}
\end{figure}

The detailed balance condition in the Fokker-Planck framework can be rewritten, leading not only to a condition that can be easily checked analytically but also to deeper insights into the fundamental causes of TRSB in NESSs. 
In fact, $J=0$ can be rewritten using the equation above:
\begin{equation}
\label{eq:rotor}
   \nabla \times \mathscr{F} = 0,
\end{equation}
where the so-called thermodynamic force $\mathscr{F}$ is given by
\begin{equation}
    \mathscr{F} = \mathcal{D}^{-1}(\mathcal{F}-\nabla \cdot \mathcal{D}).
\end{equation}

From direct computation of Eq.~\eqref{eq:rotor} one clearly sees that detailed balance is broken as soon as $\Gamma \neq0$. 
It is interesting to look at the curl of the termodynamic force $\nabla \times \mathscr{F}$ for different $\Gamma$s shown in  Fig.~\ref{fig:curl}. Comparing these plots with the movement of the net current in Fig.~\ref{fig:regions} recognizes that the regions that correspond to positive or negative $\nabla \times\mathscr{F}$ correspond to regions where the vorticity is counter-clock or anti-counterclock wise.
Therefore, the curl of the thermodynamic force plays the role of the electric density current in magnetostatic, where it induces the magnetic field. Here $\nabla \times \mathscr{F}$ is the source of the NESS~\cite{chou2011non,fang2019nonequilibrium}.
A succinct way to rephrase the above intuition is that NESSs are related to a topological symmetry breaking. 

Since we cannot construct easily the steady-state distribution due to the absence of detailed balance for $\Gamma \neq 0$, determining $P_*(\hat R)$ for generic $\Gamma$ values remains a challenge. 
In the following, we show how interesting insights can still be garnered from the steady-state PDF of the beliefs difference $\hat R^A_t-\hat R^B_t$.

\subsection{Emergent risk-aversion as termophoresis}
\label{sec:potential} 

At first glance, the use of point estimates in the update equation for the beliefs given by Eqs.~\eqref{eq:RW+f} appears overly simplistic, especially when considering its lack of direct reference to well-documented human behavioral tendencies, such as risk aversion. Risk-averse individuals demonstrate a preference, if everything else equal, for less variable options, reducing the associated risk.
However, early numerical analysis on related models revealed that the use of point estimates in the update equation of the beliefs does not neglect these tendencies, i.e., risk aversion is an emerging property of the beliefs dynamics~\cite{march1996learning}. Here we show that the Fokker-Planck framework allows us to derive this result explicitly.

In our model, the noise space dependency is solely on the difference $\hat R^A_t-\hat R^B_t$ (see Eq.~\eqref{eq:meana}). To exploit this inherent symmetry, let us introduce the coordinate transformation $(\hat R^A_t, \hat R^B_t) \rightarrow  (\hat R^A_t+\hat R^B_t, \hat R^A_t-\hat R^B_t)$
and similarly for the rewards  $(R^A_t,R^B_t)$.
Of particular interest is the observation that the update equation for $\delta\hat R_t = \hat R^A_t-\hat R^B_t$ remains independent of the variable $\hat R^A_t + \hat R^B_t$, thus implying that the detailed balance for $\delta\hat R$ holds.

The TRS of $\delta\hat R_t$ in the steady state allows for an analysis of the associated Fokker-Planck equation. In particular, the thermodynamic force $ \tilde{\mathscr{F}} =  \tilde{\mathscr{F}}[\delta \hat R]$ is given by
\begin{equation}   
    \label{eq:thermo_force}
 \tilde{\mathscr{F}} \sim \dfrac{1}{\beta}\cfrac{  -  2\delta \hat R +  \langle \delta R\rangle +   \langle R\rangle  \tanh [\Gamma  \delta \hat R]}{  \sigma^2_A a^2+\sigma^2_B (1-a)^2},    
\end{equation}
where for conciseness we haven't reported the second subleading term ($\nabla \cdot \mathcal{D}/\mathcal{D}$) and $a$ is the fraction of endowments invested in arm $A$ according to Eq.~\eqref{eq:meana}, which depends on $\delta\hat R$.
From the equation above it is clear that for intermediate $\Gamma$s, the multiplicative noise implies risk aversion: in fact, the denominator is smaller in the case of $\delta \hat R>0$ for $\sigma^2_A<\sigma^2_B$; this implies a stronger thermodynamic force towards region with $ \delta\hat R_t>0$, i.e., to belief states where the agent invests mostly on the less variable arm $A$.

Interestingly for the present discussion, the form of detailed balance given by Eq.~\eqref{eq:rotor} is known as potential condition~\cite{gardiner1985handbook}. The reason is apparent for the dynamics of $\hat R^A-\hat R^B_t$ we are discussing. In fact, one has ${P}_*[\delta\hat R] \propto \exp[\int \tilde{\mathscr{F}} ] = \exp[-\tilde{\Phi} ]$, i.e., since $\tilde{ \mathscr{F}}$ is curl-free then the thermodynamic potential $\tilde{\Phi}$ can be constructed by a simple integration of the thermodynamic force from which the standard Gibbs distribution for the associated ESS follows.

\begin{figure}[t!]
    \centering
    \hspace{-4.6cm}\includegraphics[scale = 0.5]{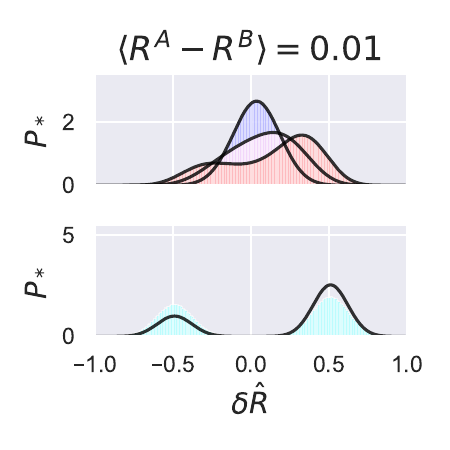}\includegraphics[scale = 0.5]{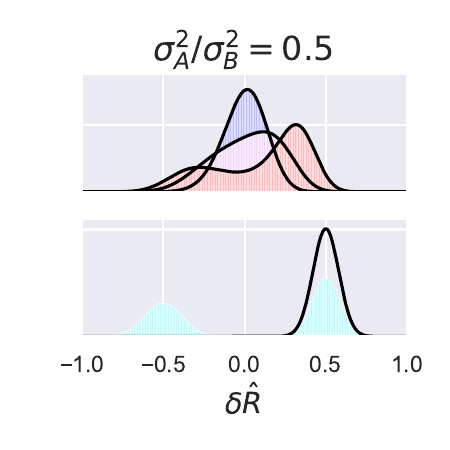}
    \caption{Comparison of stationary PDF computed from the analytical prediction (black solid lines) with the one computed from numerical simulations (colored histograms). (Left) Bandits with symmetric variances and asymmetric rewards:  $\langle R^A \rangle = 0.51$ and $\langle R^B \rangle = 0.49$. (Right) Bandits with symmetric rewards and asymmetric variances: $\sigma^2_A = \sigma^2_B/2$. The total simulation time is $10^4$ and we retain only the second half of the trajectories. }
    \label{fig:outofstationarity}
\end{figure}

The emerging risk-aversion can be visually appreciated in the plots on the right of Fig.~\ref{fig:outofstationarity}, where we compare ${P}_*[\delta\hat R]$ obtained from simulations and the one predicted from the theoretical argument above. The top two plots are obtained with parameters $\beta = 0.1, \Gamma \in {1.5,2,2.4}$, and show no difference between theory and simulation. The plots on the bottom are instead devoted to showing a  numerical issue for large $\Gamma$s ($\Gamma = 10$): although the $\tanh$ in Eq.~\eqref{eq:meana}  guarantees a unique steady state, reaching it might be numerically prohibitive. 

Notably, the way in which we recover emergent risk-aversion is exactly in line with how standard thermophoresis~\cite{busiello2023emergent}, i.e., the particles' tendency to move to cooler regions in a solution with a non-vanishing temperature gradient, arises in physical systems.

Let us remark here that not only it is possible to compute analytically ${P}_*[\delta\hat R]$,  but also the steady state PDF related to the average cumulated earned reward, represented by $R^A_t a_t + R^B_t (1-a_t)$. In fact, the cumulated earned reward at time $t$ is governed by the difference in beliefs (see Eq.~\eqref{eq:meana}). 
This leads to an interesting insight, anticipated without proof in section~\ref{sec:0}: an irreversible sequence of belief updates may -and do, in the present model- generate a time-reversible sequence of actions; furthermore, in the case of a fixed environment like the one of the present setup, also the sequence of cumulated earned rewards is time-reversible.

\section{Time irreversibility in beliefs dynamics}
\label{sec:methods}

This section is divided into two parts. First, we use Landauer's principle to define what we call cognitive energy cost and then we argue by means of theoretical arguments that it is optimized in the steady state. Finally, numerical analysis relates the time-irreversibility in the steady state to the exploration-exploitation trade-off in the modified forgetting Q-learning model in the continuous-time limit given by Eqs.~\eqref{eq:cont_lim}.

\setcounter{equation}{10}  
\subsection{Irreversibility as cognitive energy cost}

The fundamental discovery encapsulated in Landauer's principle is that the average work dissipated by an actual machine in order to make the shift from $\hat R_t$ to $\hat R_{t+1}$ is bounded from below by the irreversibility rate $\Phi$ in units of $k T$, where $k$ is the Boltzmann constant and $T$ is the temperature of the room in which the system performing this operation is working. Below we detail how this statement can be formally established. This will lead naturally to the notion of cognitive energy cost we use in this discussion.

The irreversibility rate $\Phi$ is defined as the  Kullback-Leibler divergence between the probability of observing a jump and its time reversed~\cite{seifert2012stochastic,roldan2013facultad}, i.e.
\begin{equation}
\label{eq:irr}
{\Phi_t} = D^{KL}\left[{P}_t[\left. \hat R_t\rightarrow \hat R_{t+1}] \ \right|{P}_t[\hat R_{t+1}\rightarrow \hat R_{t}] \right],
\end{equation}
where $D^{KL}[P|Q] = \int_x P(x) \log{P(x)/Q(x)}$. 
This divergence is appropriate for Markovian processes (like the one we are considering in this work)~\footnote{We note that Eq.~\eqref{eq:irr} can be derived by a more general formulation valid for non-Markovian processes, where the primary object is the Kullback-Leibler divergence between the probability of an actual path and its time-reversed twin~\cite{roldan2010estimating}. }. Let us note that the Kullback-Leibler divergence is non-negative by construction and invariant by a homogeneous dilation of the state space.

The irreversibility rate can be exactly computed in the continuous-time limit for systems described by Langevin equations like Eqs.~\eqref{eq:langevin} by means of path integrals techniques~\cite{cugliandolo2017rules, busiello2023emergent}. One obtains 
 \begin{equation}
 \label{eq:S_Flux}
\Phi_t =  \langle  v_t \cdot \mathscr{F}_t \rangle,
 \end{equation}
 where $v_t = J_t/P_t$ is the net directed velocity of the beliefs in the 2-dimensional space $\hat R$ and $\langle \cdot \rangle$ stands for the average over $P_t$. Therefore, $\Phi$ is the dissipated power from the thermodynamic force $\mathscr{F}$ in units of $k T$. Hence, we identify the irreversibility rate $\Phi_t$ with the fundamental cognitive energy cost needed in order to perform a shift from $\hat R_t$ to $\hat R_{t+1}$.

 Let us recover a previous result anticipated in Sec.~\ref{sec:FP}, related to the fact that the NESS is generated by $\nabla \times \mathscr{F}$. Given the new quantity $\Phi$ we have introduced, this means that $\Phi_* \neq 0$ for $\Gamma \neq 0$. This result can be recovered as follows. In the steady state, the velocity follows circulating lines (see the currents in Fig.~\ref{fig:regions} again and remember that $v_t = J_t/P_t$). One can calculate the average over the whole state space in Eq.~\eqref{eq:S_Flux} as an average over these closed lines. The dissipated power by the thermodynamic force on a closed loop is in general positive in the steady state for $\Gamma \neq 0$ because, by applying Stokes' theorem, the line integral receives a contribution from the surface integral of $\nabla \times \mathscr{F}$, which we know from previous analysis being general different from zero (see Fig.~\ref{fig:curl}). 

Equation~\eqref{eq:S_Flux} gives another interesting insight: in the steady state the velocity has to be aligned to the non-conservative part of $\mathscr{F}$ since we know that $\Phi_t$ is non-negative by construction. App.~\ref{eq:Lyap} will prove that actually in the steady state, the velocity is maximally aligned with the non-conservative thermodynamic force compatibly with a minimal dissipation along closed lines.

\subsection{Numerical results}
\label{sec:2}

We focus on the analysis of the irreversibility rate in the steady state of the belief dynamics given by Eqs.~\eqref{eq:cont_lim}. For fixed bandit configuration, the only interesting dynamics in the continuous-time limit is the one for fixed $\beta$ and varying $\Gamma$s. 

In fact, due to dimensional analysis considerations, if we let vary $\beta$ for fixed $\Gamma$s, the irreversibility rate will simply scale as $\beta$. This means that the dynamics is exactly analogous for different $\beta$s, the only thing that changes is the typical recurrence time, which scales as $\sim 1/\beta$. I.e., for decreasing $\beta$, the recurrence time will increase, and therefore the irreversibility rate will diminish.

We consider three different scenarios: the case of completely symmetric arms (like the one discussed in Fig.~\ref{fig:regions} and \ref{fig:curl}), the case of asymmetric average rewards, and finally the case with asymmetric variances (respectively shown already in the left and right plots of Fig.~\ref{fig:outofstationarity}). 

\begin{figure}[t!]
    \centering
    \hspace{-9cm}\includegraphics[scale =0.59]{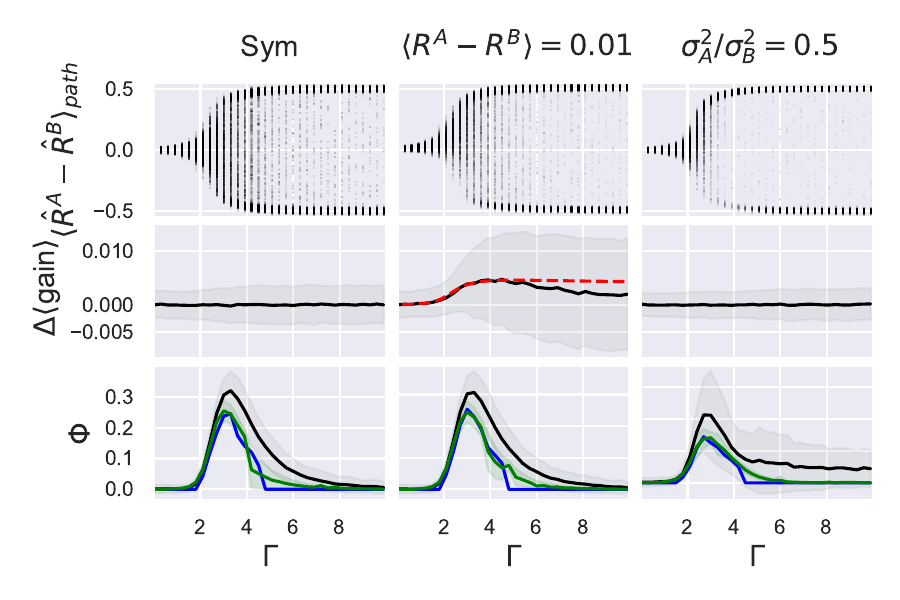}
    \caption{Numerical results for different scenarios and for different $\Gamma$s. (Left) Symmetric bandits; (Center) Asymmetric bandits in the average reward. (Right) Asymmetric bandits in the variance of the rewards. Displayed metrics include (from top to bottom) average difference in beliefs, average earned reward minus the one related to $\Gamma = 0$, and irreversibility rate. The red line in the central plot is related to the theoretical value of this metric. Black, blue and green lines in the lower panel represent $\Phi$ calculated from Monte Carlo simulations, a Neural Network and a Gradient Boosting approach, respectively. These estimators are based on exactly the same set of trajectories.}
    \label{fig:enter-label}
\end{figure}

For each scenario, three metrics are exhibited in Fig.~\ref{fig:enter-label}. Note that in order not to incur in degenerate diffusion matrix in the case of large $\Gamma$s, we add a small exogenous noise to the update equations (see App.~\ref{sec:app}).

\begin{itemize}[leftmargin=*,align=left]
    \item[Average difference in beliefs:] each point corresponds to the average difference of belief of each trajectory. 

    By looking at this metric one can again see that at high exploitation levels trapping states emerge. Moreover, this metric gives a clear picture of the average fraction of time passed in a given belief state.
\end{itemize}
    
\begin{itemize}[leftmargin=*,align=left]
    \item [Average earned reward:] Each point corresponds to the average earned reward across the trajectories minus the one obtained with passive learning, i.e.,  $\Gamma = 0$. The red line is obtained analytically starting from Eq.~\eqref{eq:thermo_force} (see the discussion at the end of sec.~\ref{sec:potential}). The quantitative disagreement at large $\Gamma$s is due to the fact that the equilibration time exceeds the simulation time for large $\Gamma$s (see bottom plots in Fig.~\ref{fig:outofstationarity} and the discussion below them). 

    This metric reflects the mean cumulative reward earned $R^A_t a_t + R^B_t (1-a_t)$ by the agent, thereby quantifying the system's operational efficiency. Note that in the case where the arms have different average rewards (central plots), the maximum average earned reward is obtained for moderate $\Gamma$s. In the third scenario, instead, the largest fraction of time passed in the less volatile arm is obtained again for moderate $\Gamma$s.
    \end{itemize}

\begin{itemize}[leftmargin=*,align=left]
    \item [Irreversibility rate:] each point corresponds to the average irreversibility rate across the trajectories.

In order to compute the irreversibility rate numerically from Monte Carlo simulations, we note that  Eq.~\eqref{eq:S_Flux} after an integration by parts, leads to~\footnote{Interestingly, the irreversibility rate is composed by two terms: the first corresponds to exploration, while the second is associated with the contraction of the belief space due to forgetting~\cite{contraction_phase_space}.} 
\begin{equation}
\begin{split}
\label{eq:equalityphipi}
    \Phi_t =& \langle (\mathcal{F}_t-\nabla \cdot \mathcal{D}_t)^\mathsf{T}\mathcal{D}_t^{-1}(\mathcal{F}_t-\nabla \cdot \mathcal{D}_t)\rangle 
    \\ 
    &- \langle \nabla \cdot (\mathcal{F}_t-\nabla \cdot \mathcal{D}_t) \rangle.
\end{split}
\end{equation}    
$\mathcal{F}_t$ and $\mathcal{D}_t$ are derived explictely in Eq.~\eqref{eq:F} and Eq.~\eqref{eq:D} respectively.

    On top of the black line provided by Eq.~\eqref{eq:equalityphipi}, two additional benchmarks calculated directly from Eq.~\eqref{eq:irr} are presented: the blue line is based on a recently proposed Neural Network approach~\cite{kim2020learning}, while the green line is provided by an algorithm that maps the problem of calculating the irreversibility rate onto a classification task~\cite{seif2021machine}, by leveraging on gradient boosting techniques. Crucially, these additional estimators do not need any information about the model except the tacitly assumed Markovian property by using Eq.~\eqref{eq:irr} of the underlying process. The reason why the Monte Carlo estimator is consistently above the others is related to the fact that no spatial coarse-graining is applied in this case since full knowledge of the underlying model is provided.
    \end{itemize}

 The irreversibility rate $\Phi$ is null at both exploitation parameter extremities, in sync with previous analyses done in this paper. A noteworthy crest is observed at median exploitation parameters; this indicates a belief dynamics propitious to humans in asymmetric bandit scenarios, being comfortably distant from bifurcation-prone zones. In fact, by looking at the top and center plots in the asymmetric bandit scenarios, one can see that the exploitation level related to the maximal irreversibility rate corresponds to a heightened average earned reward variance and lowered average earned reward variability, respectively.

\section{Discussion}
\label{sec:3}

We linked the irreversibility rate associated with beliefs dynamics to a thermodynamically consistent measure of cognitive energy cost, according to Landauer's principle.

This idea has been applied to explore the role of time-reversal symmetry breaking in a simple but paradigmatic setup: we modified a standard prediction error-based beliefs dynamics~\cite{Katahira} to account for finite memory, exploitative behavior and limited resources within a two-armed bandit problem.

First, we provide a mapping of the decision-making model onto a model for active particles, i.e., particles able to spend energy to move. A side result is the formal identification of emerging risk-aversion of beliefs dynamics in the present setup and standard thermophoresis.

 The combination of theoretical and numerical analysis has shown that intermediate exploitative behavior produces maximum -yet efficient- cognitive energy cost as well as the best trade-off between exploration and exploitation. 
 
 Therefore, this stylized model suggests a plausible evolutionary mechanism that underscores the likelihood of biological entities to be optimized to function in maximally out-of-equilibrium states~\cite{inzlicht2018effort}. This insight is in line with Prigogine’s principles of the natural emergence of optimal maximally 
dissipative structures~\cite{prigogine1977nobel}.

Below we will illustrate a number of model-dependent and model-free future research directions.

 The present model can be modified to account for  positivity or confirmation biases, and this will likely yield to more exotic thermoporetic effects~\cite{palminteri2022choice,chambon2020information,sugawara2021dissociation}: in fact, positivity bias is known to lead to emergent risk-seeking behavior; the finding of this paper suggest that the Fokker-Planck description should instead lead to negative thermophoresis, where a solute moves from cooler to hotter regions~\cite{liu2023negative}, 

Another interesting modification is to consider different decision-making rules. For instance, scale-invariant~\cite{chater1999scale,dehaene2003neural,bavard2018reference,bavard2023functional} decision-dynamics, which, by continually relating the difference in beliefs to a shifting reference level, leads to an inherently adaptive even in the case of evolving environments. Another modification in the decision dynamics is to introduce some inertia~\cite{moran2020force,palminteri2023choice}, leading to a new source of exploitative behavior.

The cognitive energy costs incurred during transitions between states due to environmental shifts can be analyzed. This will likely yield insights into cognitive plasticity and its interplay with cognitive energy cost and time irreversibility. This analysis could shed light on the emergence of aforementioned cognitive biases as an effective way of reducing cognitive energy cost in changing environments.

A more fundamental question is related to the fact that the separate retention of subjective belief for each arm, substantiated by research in neuronal bases of decision-making, is essential to the present discussion: we proved in fact that the decision-maker could reach the same rewards if he retained only information about the difference in belief; this seems at first beneficial because it allows the decision maker to not incur in any cognitive energy cost in the steady state but raises an important question devoted to future research: what could be an analog of a `no free-lunch theorem' in the present setting? Can we relate the intrinsic cognitive energy cost to retain separate beliefs to some objective potential benefit in cases where additional options become available over time?

On the other hand, subjective belief dynamics are not solely of interest to cognitive scientists. In this regard, a  more applied research question comes from the following consideration: a large amount of person-specific data available from social platforms already allows us to compute proxies for subjective beliefs, such as political leaning~\cite{cinelli2021echo}. We believe that an analysis of the time irreversibility in the subjective belief dynamics of single individuals in social networks can shed light on very imminent questions such as `are social networks responsible for heightened levels of polarization in our societies?'.

In conclusion, the present analysis shows how out-of-equilibrium physics breakthroughs can help to decipher the underlying reason why cognitive systems navigate and adapt to their continuously evolving belief landscapes in the way they do.

\section*{Acknowledgements}
I am indebted to Christian Bongiorno for his indispensable assistance with the numerical aspects of this research and to Stefano Palminteri for directing us to relevant literature in the cognitive neuroscience domain. The discussion was significantly enriched by Damien Challet's broader perspectives on the topic. I also acknowledge fruitful discussions with Cristiano Pacini, Matteo Marsili, Massimo Vergassola, Stefano Celani, Edgar Roldan, Matteo Sireci, Daniel Busiello, and Walter Quattrociocchi. The author is supported by the Agence National de la Recherche (CogFinAgent: ANR-21-CE23-0002-02).

\appendix

\renewcommand{\theequation}{A.\arabic{equation}}
\setcounter{equation}{0}  

 \section{Analysis of Lyapunov function}
 \label{eq:Lyap}
In order to have insights about how $\Phi$ is optimized as the NESS is reached, it is useful to study a particular Lyapunov function of the dynamics. A Lyapunov function is such that its temporal derivative is always non-positive, meaning that its fixed point corresponds to the steady state of the dynamics.

Consider the function $\mathcal{L}$ given by
\begin{equation}
\label{eq:luapunov}
\mathcal{L} = D^{KL}\left[P_t| P_*\right].
\end{equation}
By taking the time derivative of $\mathcal{L}$ and  inserting the Fokker-Planck equation one obtains:
\begin{eqnarray}
\label{eq:14}
    \frac{d\mathcal{L}_t}{dt} &=& - \Pi_t + \langle v_t \mathcal{D}^{-1} v_*\rangle 
    \\
    \label{eq:15}
    &=& -\langle (v_t -v_*)\mathcal{D}_t^{-1}(v_t -v_*)\rangle \leq 0,
\end{eqnarray}
where $\Pi_t = \langle v_t \mathcal{D}^{-1} v_t\rangle$ in the first equality is the so-called entropy production in the stochastic thermodynamics literature~\cite{seifert2005entropy,tome2006entropy,seifert2012stochastic,busiello2019entropy}. The second equality can be established by noting that~\cite{van2010three} $\langle v_t \mathcal{D}^{-1} v_* \rangle = \langle v_* \mathcal{D}^{-1} v_*\rangle$. The final inequality in Eq.~\eqref{eq:15}, trivially follows since the final term is quadratic in $v_t-v_*$ and $\mathcal{D}$ is semi-positive definite by construction. This proves that $\mathcal{L}_t$ is a Lyapunov function of the dynamics.

Let us make an important remark: $\Pi_t \geq 0$ by definition because it is quadratic in the thermodynamic velocities $v_t$ and inversely proportional to the diffusion matrix, which is semi-positive definite by construction. In ESSs, $\Pi = 0$ because $v = 0$ by definition; therefore, $\Pi >0$, i.e., the case where currents are present, is a clear marker of irreversible dynamics.

Following a similar reasoning, one can see that in Eq.~\eqref{eq:14} the negative time derivative of the Lyapunov function has been written as the sum of a non-positive and a non-negative term (remember that $\langle v \mathcal{D}^{-1} v_* \rangle = \langle v_* \mathcal{D}^{-1} v_*\rangle$), suggesting that in the vicinity of the steady state, the first is maximized and the second is minimized. 

Interestingly, the second term in Eq.~\eqref{eq:14} can be rewritten by simply using the identity $v_* = J_*/P_*$ and the definition of $J$ given by Eq.~\eqref{eq:current}. One obtains
\begin{equation}
\label{eq:16}
    \langle v \mathcal{D}^{-1} v_*\rangle = \Phi_t -\langle v \cdot \nabla \log[P_*] \rangle.
\end{equation}
In the steady state, the second term in the r.h.s. of the equation above can be rewritten after a partial integration as $\langle \nabla \cdot v_* \rangle_*$, where $\langle \cdot \rangle_*$ indicates an average over the steady state PDF $P_*$; this term has to be zero in a NESS with a compact state space since the occupied state space in the steady state is no longer contracting or expanding.
Therefore, along the dynamics, $d \mathcal{L}_t/dt$ goes to zero by minimizing the entropy production $\Pi_t = \langle v_t \mathcal{D}^{-1} v_t \rangle$ while maximizing the dissipation of the thermodynamic force along the closed lines created in the vicinity of the steady state by probability currents. 

From Eq.~\eqref{eq:14} evaluated in the steady state one obtains the well-known result $\Phi_* = \Pi_*$, i.e., in the steady state the irreversibility rate, also known as entropy flux, is equal to the entropy production.
The equation $\Phi_* = \Pi_*$ can be interpreted as a form of energy conservation, echoing the interpretation in physics. In fact, the entropy flux is the average dissipated power in units of $kT$ done by the thermodynamic force $\mathscr{F}$, as previously emphasized. On the other hand, $\Pi$ is analogous to the kinetic energy of the active particle with velocity field $v_t$ and mass  $\mathcal{D}^{-1}$; this is tantamount to saying that the inertia of the particle is lower in a noisier environment. 

Let us recapitulate what we have obtained.

The main result of this section is that the combination of Eq.~\eqref{eq:14},\eqref{eq:15} and \eqref{eq:16} implies that the NESS is the least dissipative state compatible with a velocity that is maximally aligned with the non-conservative part of thermodynamic force $\mathscr{F}$, therefore suggesting an efficient (thermodynamically speaking) information processing in the steady state~\cite{busiello2023emergent}.

\renewcommand{\theequation}{B.\arabic{equation}}
\setcounter{equation}{0}

\section{Model used for simulations}
\label{sec:app}
The model for which we are going to investigate quantitatively $\Phi_*$ is given by:
\begin{equation}
\label{eq:cont_lim_fin}
\begin{split}
\cfrac{d \hat R^A_{t}}{dt} &= -\beta\left(  \hat R^A_t +  a_t R^A_t+\eta^A_t\right)
\\
\cfrac{d \hat R^B_{t}}{dt} &=  -\beta\left( \hat R^A_t + (1-a_t) R^B_t+\eta^B_t\right)
\end{split}    
\end{equation}
where we made one modification with respect to Eqs.~\eqref{eq:cont_lim}: 
we added two small exogenous white noises, $\eta^A_t$ and $\eta^B_t$, which are needed in order to have a well-defined two-dimensional diffusion matrix in the large-$\Gamma$ region, where, in the absence of such noises, it would become a singular matrix. I set the variances of $\eta^A_t$ and $\eta^B_t$ so that $\text{var}[\eta^A] = \text{var}(\eta^B) = \sigma^2_\eta 
 \ll \sigma^2_{A},\sigma^2_{B}$.

The derivation of the Fokker-Planck equation~(see Sec.~\ref{sec:fokker_planck}) leads to:
\begin{equation}
\label{eq:F}
    \mathcal{F}_t = \beta \begin{bmatrix}
        -  \hat R^A_t +   a_t \langle R^A \rangle 
        \\
        -  \hat R^B_t +  (1-a_t) \langle R^B \rangle 
    \end{bmatrix}
\end{equation}
and
\begin{equation}
\label{eq:D}
    \mathcal{D}_t =(\beta/2)^2 \begin{bmatrix}
        \sigma^2_A a_t^2 + \sigma^2_\eta, & 0
        \\
        0, & \sigma^2_B (1-a_t)^2 + \sigma^2_\eta
    \end{bmatrix}
\end{equation}    
These are the expressions of $\mathcal{F}$ and $\mathcal{D}$ we use to quantify the irreversibility rate from Monte Carlo simulations by means of Eq.~\eqref{eq:equalityphipi} in Fig.~\ref{fig:enter-label}.

\bibliography{biblio}
\end{document}

%% file: preamble.tex
\AtBeginDocument{\pagenumbering{arabic}}
\usepackage{adjustbox}
\usepackage[]{graphicx}
\usepackage{wrapfig}
\usepackage[caption=false]{subfig}
\usepackage{color}
\usepackage{wrapfig}
\usepackage{mathrsfs}
\usepackage{xcolor}
\usepackage[titletoc,toc]{appendix}
\usepackage{graphbox}
\usepackage{enumitem}

\usepackage{romannum}

\usepackage[colorlinks,bookmarks=false,urlcolor=blue, citecolor=blue, linkcolor = blue]{hyperref}

\usepackage{tikz}
\usetikzlibrary{positioning,arrows}
\usetikzlibrary{decorations.pathmorphing}
\usetikzlibrary{decorations.markings}
\usetikzlibrary{calc}
\usetikzlibrary{patterns}
\usetikzlibrary{shapes.misc}
\tikzset{
phys/.style={thick, postaction={decorate}, decoration={markings, mark=at position .7 with {\arrow[]{triangle 45}}}},
res/.style={thick, decorate, decoration={snake,segment length=7, amplitude=1.5}},
arrow/.style={thick, draw=white, postaction={decorate}, decoration={markings, mark=at position .6 with {\arrow[black]{triangle 45}}}}
 }
 
\usepackage{hyperref}
\hypersetup{
    colorlinks=true,
    linkcolor=black,
    citecolor=black,
    filecolor=black,
    urlcolor=black
}

\usepackage{etoolbox}
\patchcmd{\thebibliography}{%
  \section*{\refname}%
}{%
  \section*{\refname}%
  \small
}{}{}
\renewcommand{\refname}{References}

\usepackage[absolute]{textpos}
\usepackage[bottom]{footmisc}